\documentclass[a4paper,11pt]{article}

\usepackage[utf8]{inputenc}
\usepackage[T1,T2A]{fontenc}
\usepackage{amsmath,amsfonts,amssymb}

\usepackage{cite}

\usepackage{xcolor}

\usepackage{graphicx}
\usepackage{wrapfig}

\usepackage{geometry}
\newcommand{\lb}[0]{\left(}
\newcommand{\rb}[0]{\right)}

\newcommand{\pz}{\partial_z}

\begin{document}

\renewcommand*{\thefootnote}{\fnsymbol{footnote}}

\begin{center}
{\Large\bf Ultraviolet regularization of energy between two static sources in the bottom-up holographic approach to strong interactions}
\end{center}
\bigskip
\begin{center}
{S.S. Afonin\footnote{E-mail: \texttt{s.afonin@spbu.ru}.}
}
\end{center}

\renewcommand*{\thefootnote}{\arabic{footnote}}
\setcounter{footnote}{0}

\begin{center}
{\small\it Saint Petersburg State University, 7/9 Universitetskaya nab.,
St.Petersburg, 199034, Russia}
\end{center}

\bigskip

\begin{abstract}
It is well known that the potential energy between two heavy quarks carries an important information about the physics of confinement.
Using the Wilson loop confinement criterion and the Nambu-Goto string action, this energy can be derived within the bottom-up holographic
approach to strong interactions. We recapitulate the standard holographic derivation of the potential between two static sources with
emphasis on the physical interpretation of the results. After that we consider the problem of regularization of arising
ultraviolet divergence in a general case. Here the term ``ultraviolet'' means small values of the holographic coordinate which is related
with the inverse energy scale in the holographic duality. We show that in the case of widely used Soft-Wall holographic models, in principle,
many ultraviolet divergences may appear, although in practice the appearance of more than two different divergences looks somewhat exotic.
Some possible subtraction schemes are discussed. Different schemes lead to a different constant shift of the potential energy and
this entails a certain scheme-dependence of holographic predictions for constant term in resulting Cornell-like confinement potentials.
\end{abstract}





\section{Introduction}

The AdS/CFT correspondence in string theory~\cite{mald,witten,gub} inspired the emergence of holographic
approach to QCD. The corresponding model building has two vast branches --- the top-down (where one starts from some
brane constructions in the string theory) and bottom-up (that starts from the real QCD) holographic
approaches. Both directions of research have led to the construction of many working phenomenological models. The second direction,
however, became much more popular in the QCD phenomenology. The bottom-up holographic models have enjoyed a considerable phenomenological success
in applications to various problems in the physics of strong interactions. Perhaps the most widely used class of such models
is the class of the so-called Soft-Wall (SW) holographic models originally introduced in~\cite{son2,andreev}.
The SW models were successfully applied to the description of hadron Regge spectroscopy, hadron form-factors, QCD thermodynamics,
and other phenomenology related to the non-perturbative strong interactions (see, e.g.,
a review of recent literature in~\cite{Afonin:2021cwo}).
The simplest SW models are the bottom-up holographic models with a certain static
gravitational background that reproduces the Regge spectrum of light mesons.
In most cases, this background is not considered as a solution of some known dual 5D gravitational theory,
rather it interpolates in a model way some important contributions from a hypothetical unknown dual
string theory for QCD. A conceptual reason seems to lie in the fact that QCD is weakly coupled at high energy due to
asymptotic freedom and the dual description of weakly coupled theories requires highly curved space-times
(since the AdS/CFT correspondence is a strong-weak duality)
for which the approximation by usual general relativity is not enough to analyze the dual theory.
As the corresponding dual formulation cannot be given in terms of a gravitational field theory one must involve,
strictly speaking, the full dual string theory~\cite{PhysToday}. This task is too difficult to implement\footnote{See,
however,~\cite{Iatrakis:2010jb} and references therein where it was advocated that the structure of
holographic theories for QCD should arise within the framework of some
non-critical string theory, in which the condensation of scalar tachyon can naturally generate
the background of SW model; actually this hypothesis was suggested already in the pioneering paper~\cite{son2}.}
but phenomenologically one can guess a simple background that correctly reproduces the high-energy behavior of
two-point correlation functions in QCD together with the leading power-like non-perturbative corrections
known from the QCD sum rules.
And just such kind of background was proposed in papers~\cite{son2,andreev} which gave rise to the SW holographic approach.
The Regge form of the spectrum, which serves usually as the main motivation for the SW holographic models, in fact, is a
consequence of the correct analytical properties of the correlation functions in the SW models~\cite{Afonin:2010fr}.

Among numerous phenomenological applications of SW model
there was a derivation of heavy-quark potential~\cite{Andreev:2006ct}.
The found potential turned out to be close to the well-known Cornell potential~\cite{bali}.
Similar results for the confinement potential were also obtained in many other bottom-up
holographic models, see, e.g.,~\cite{heavy} and references therein.
In the process of holographic calculation of static energy, an ultraviolet divergence arises
which must be regulated. In the present note, we analyze the ultraviolet
renormalization of potential energy between two heavy sources in a general case
of bottom-up holographic models.

The paper is organized as follows.
The basics of SW holographic model are introduced in Section~2. In Section~3,
the holographic derivation of the potential between two static sources is explained. Then, in Section~4,
we consider the problem of regularization of arising ultraviolet divergences in a general case.
Our summary is given in Section~5.

\section{Soft Wall holographic model}

The action of SW holographic model can be defined as~\cite{andreev,Afonin:2021cwo}
\begin{equation}
\label{sw}
  S=\int d^4\!x\,dz\sqrt{G}\,\mathcal{L},
\end{equation}
where $G=|\text{det}G_{MN}|$, $\mathcal{L}$ is a Lagrangian density of some
free fields in a modified 5D Anti-de Sitter (AdS\(_5\)) space (modified in the infrared region)
which, by assumption, are dual on the AdS\(_5\) boundary to some QCD operators
in the spirit of the AdS/CFT correspondence~\cite{mald,witten,gub}.
The metric is given by a modification of the Poincar\'{e} patch of the AdS$_5$ space,
\begin{equation}
\label{az_metric}
ds^2=G_{MN}dx^M\!dx^N=h\frac{R^2}{z^2}\lb\eta_{\mu\nu}dx^\mu\! dx^\nu-dz^2\rb,
\end{equation}
\begin{equation}
\label{h}
  h=h(z),\qquad h(0)=1.
\end{equation}
Here $\eta_{\mu\nu}=\text{diag}\lbrace1,-1,-1,-1\rbrace$, $z>0$ is the holographic coordinate,
$R$ denotes the radius of AdS$_5$ space. The holographic coordinate $z$ has the physical meaning
of inverse energy scale~\cite{mald,witten,gub}, so the ultraviolet limit in the parametrization~\eqref{az_metric}
is the limit $z\rightarrow0$ which gives in~\eqref{az_metric} the 4D Minkowski space as a boundary of the AdS$_5$ space.
The function $h(z)$ defines a concrete model and introduces a mass scale into the model.
For instance, $h(z)= e^{cz^2}$ in the original SW model~\cite{andreev}, where $c$ represents
a parameter of dimension of mass squared that becomes proportional to the slope of Regge mass spectrum.
As was mentioned above, the standard SW model is defined in the probe
approximation, i.e., the metric is not backreacted by matter fields in $\mathcal{L}$
--- such backreaction is assumed to be suppressed in the large-$N_c$ limit (it should be recalled that,
strictly speaking, the holographic approach is formulated in this limit only) or to be effectively
taken into account by a choice of function $h(z)$.
The simplicity of the probe approximation provides an analytical control at each stage of calculations.

In the case of massless 5D fields, the model~\eqref{sw} can be rewritten in an equivalent form~\cite{son2,Afonin:2021cwo},
\begin{equation}
\label{sw2}
  S=\int d^4\!x\,dz\sqrt{G}\,\tilde{h}\mathcal{L},
\end{equation}
in which the metric modification~\eqref{h} is set to $h=1$ and replaced instead by the ``dilaton'' background $\tilde{h}$ in the action.
The formulation~\eqref{sw2} is the most popular in the literature, with various Lagrangians
and backgrounds (e.g., $\tilde{h}(z)=e^{\tilde{c}z^2}$ in the original SW model~\cite{son2}, the sign of mass parameter $\tilde{c}$ can be different).

\section{Holographic Wilson loop and confining behavior}

\begin{wrapfigure}{r}{0.4\textwidth}
  \vspace{-9mm}
  \begin{center}
    \includegraphics[width=0.38\textwidth]{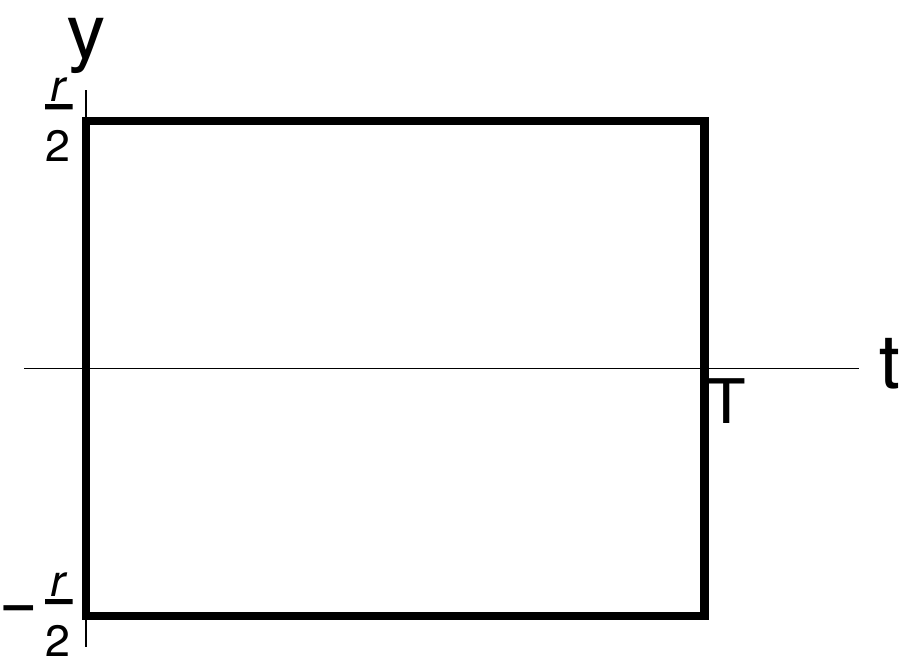}
  \end{center}
  \vspace*{-7mm}
  \caption{\small A picture of the Wilson loop considered in the text.}
  \label{wilson_loop}
\end{wrapfigure}
The holographic derivation of a static potential between two heavy quarks was
originally proposed by Maldacena in~\cite{Maldacena:1998im}.
The given derivation is based on a holographic interpretation of the expectation value of
the Wilson loop in a $SU(N)$ Yang-Mills theory,
\begin{equation}
W(\mathcal{C})=\frac{1}{N}TrPe^{\oint_\mathcal{C}\!A_\mu dx^\mu}.
\end{equation}
Namely, one considers a rectangular Wilson loop \(W(\mathcal{C})\) placed on
the 4D boundary of Euclidean 5D space with the Euclidean time coordinate \(0\leq t \leq T\),
the remaining spatial coordinate spans the interval \(-r/2 \leq y \leq r/2\), the corresponding
contour $\mathcal{C}$ is shown in Fig.~\ref{wilson_loop}.
The infinitely heavy quark and antiquark are set at $y=-r/2$ and $y=r/2$, respectively.
The Wilson loop describes the creation of a quark-antiquark pair at some time $t_1$, interaction of
the created quark and the antiquark during a period of time $T$, and the
annihilation of the pair at time $t_2=t_1+T$.
In the gauge theory,
the expectation value of this loop in the limit \(T\to\infty\) is known to be proportional to
\(\left\langle W(\mathcal{C})\right\rangle\sim e^{-TE(r)}\),
where \(E(r)\) represents the potential energy of the static quark-antiquark pair.

Based on the general prescriptions of the AdS/CFT correspondence~\cite{mald,witten,gub},
Maldacena put forward an idea that the same expectation value of Wilson loop should be equal to the partition function
of the dual open string, \(\left\langle W(\mathcal{C})\right\rangle\sim Z_\text{string}(\mathcal{C})\),
with the string worldsheet ending on the contour $\mathcal{C}$ at the boundary of AdS$_5$ space~\cite{Maldacena:1998im}.
In the saddle-point approximation, the partition function is given by the classical action,
$Z_\text{string}(\mathcal{C})\sim e^{-S}$, where \(S\) is the proper area of an open
string worldsheet which at the boundary of AdS$_5$ describes the loop $\mathcal{C}$.
Thus, in the low-energy limit of the dual string theory, it is natural to expect
\(\left\langle W(\mathcal{C})\right\rangle\sim e^{-S}\)~\cite{Maldacena:1998im}.
Then the static energy of the configuration directly follows,
\begin{equation}
\label{energy}
\left\langle W\right\rangle\sim e^{-TE(r)}\sim e^{-S}\quad \Rightarrow\quad E=\frac{S}{T}.
\end{equation}

The idea of this method was further developed in many papers within the AdS/CFT correspondence,
in particular in~\cite{Drukker:1999zq,Brandhuber:1998bs,Kinar:1998vq}.
The method was later applied to the vector SW holographic model in~\cite{Andreev:2006ct}.
The asymptotics of the obtained potential at large and small distances qualitatively
reproduced the Cornell potential,
\begin{equation}
\label{cornell}
V(r)=-\frac{\kappa}{r}+\sigma r + C.
\end{equation}
This potential was accurately measured in the lattice simulations and is now widely used in the heavy-meson spectroscopy~\cite{bali}.
The calculation of Ref.~\cite{Andreev:2006ct} was further extended to the case of
SW models generalized to arbitrary intercept parameter (i.e., when the linear Regge spectrum
has a general form $m^2_n\sim an+b$, where the intercept $b$ is arbitrary) and to the scalar SW model
in~\cite{Afonin:2021zdu,Afonin:2022aqt}. Below we recapitulate the main steps of derivation of the static energy.

In the hadronic string theory, the gluonic degrees of freedom are usually described by the Nambu-Goto action,
\begin{equation}
\label{ng}
  S=\frac{1}{2\pi\alpha'}\int d^2\xi\sqrt{\det \left[G_{MN}\partial_\alpha X^M\!\partial_\beta X^N\right]},
\end{equation}
where \(\alpha'\) is the inverse string tension, \(X^M\) are the string coordinates
functions which  map the parameter space of the
worldsheet \(\lb\xi_1,\xi_2\rb\)  into the space-time, and \(G_{MN}\) is the
metric of the bulk space.
It is therefore natural to take the action~\eqref{ng} as the worldsheet area.

The derivation of the potential energy from the Wilson loop proceeds in the Euclidean space.
The Euclidean version of the metric~\eqref{az_metric} takes the form
\begin{equation}
\label{metric}
  \text{G}_{MN}=\text{diag}\left\lbrace\frac{R^2}{z^2}h,\dots,\frac{R^2}{z^2}h\right\rbrace.
\end{equation}
Let us parametrize the string worldsheet as in Fig.~\ref{wilson_loop}, \(\xi_1=t\) and \(\xi_2=y\).
Integrating over \(t\) from \(0\) to \(T\) in~\eqref{ng}, we arrive at the action
\begin{equation}
\label{ng_action}
  S=\frac{T}{2\pi\alpha'}\int\displaylimits_{-r/2}^{r/2}dy\,\sqrt{\text{G}_{00}\text{G}_{yy}+\text{G}_{00}\text{G}_{zz}\,z'^2}
  =\frac{TR^2}{2\pi\alpha'}\int\displaylimits_{-r/2}^{r/2}dy\,\frac{h}{z^2}\sqrt{1+z'^2},
\end{equation}
where \(z'=\frac{dz}{dy}\). The obtained Lagrangian,
\begin{equation}
  \mathcal{L}=\frac{h}{z^2}\sqrt{1+z'^2},
\end{equation}
leads to the conjugate momentum,
\begin{equation}
\label{mom}
  p=\frac{\delta\mathcal{L}}{\delta z'}=\frac{hz'}{z^2\sqrt{1+z'^2}},
\end{equation}
which gives the Hamiltonian
\begin{equation}
\label{ham}
  \mathcal{H}= pz'- \mathcal{L} = -\frac{h}{z^2\sqrt{1+z'^2}}.
\end{equation}
Since $\mathcal{H}$ does not depend explicitly on $y$, its value is a constant of motion.
This conserving quantity represents the first integral of equation of motion following from
the action~\eqref{ng_action} and appears from the translational invariance of this action (there is no explicit dependence
of the Lagrangian on \(y\)). If $z(y)$ is an even function (as it will be in our case), there is a maximum value
\begin{equation}
\label{not_0}
  z_0=z(0),
\end{equation}
which is achieved at $y=0$ since at the ends of the Wilson loop we have \(z(\pm r/2)=0\) and the system
is symmetric. At that point $z'(0)=0$ and also $p=0$ in~\eqref{mom}. Therefore, the constant of motion is
\begin{equation}
  \mathcal{H}(z_0,0)= -\frac{h_0}{z_0^2},
\end{equation}
where
\begin{equation}
\label{not0}
  h_0=\left.h\right|_{z=z_0}.
\end{equation}
Now we can extract the differential equation for the geodesic line,
\begin{equation}
\label{geod}
  z'\doteq\frac{dz}{dy}=\sqrt{\frac{h^2}{h_0^2}\frac{z_0^4}{z^4}-1},
\end{equation}
and use it to write down the final expression for the distance between two static sources situated at $y=\pm r/2$ in the ordinary space,
\begin{equation}
\label{x_z_int}
  r=\int\displaylimits_{-r/2}^{r/2}dy=2\int\displaylimits_{0}^{z_0}dz\,\frac{h_0}{h}\frac{z^2}{z_0^2}
  \frac{1}{\sqrt{1-\frac{h_0^2}{h^2}\frac{z^4}{z_0^4}}}.
\end{equation}
Combining~\eqref{energy},~\eqref{ng_action}, and~\eqref{geod} we obtain the static energy,
\begin{equation}
\label{E1}
  E=\frac{R^2}{2\pi\alpha'}\int\displaylimits_{-r/2}^{r/2}dy\,
  \frac{h}{z^2}\sqrt{1+z'^2}=\frac{R^2}{\pi\alpha'}\int\displaylimits_{0}^{z_0}dz\,\frac{h}{z^2}
  \frac{1}{\sqrt{1-\frac{h_0^2}{h^2}\frac{z^4}{z_0^4}}}.
\end{equation}
This energy can be also written in the following forms
\begin{multline}
\label{E2}
  E=\frac{R^2}{2\pi\alpha'}\int\displaylimits_{-r/2}^{r/2}dy\left(
  \frac{h}{z^2}\sqrt{1+z'^2}\pm\frac{h_0}{z_0^2}\right)=\\
  \frac{R^2}{2\pi\alpha'}\left[\frac{h_0}{z_0^2}\int\displaylimits_{-r/2}^{r/2}dy+
  \int\displaylimits_{-r/2}^{r/2}dy\left(\frac{h}{z^2}\sqrt{1+z'^2}-\frac{h_0}{z_0^2}\right)\right]=\\
  \frac{R^2}{2\pi\alpha'}\left[\frac{h_0}{z_0^2}\,r+
  2\int\displaylimits_{0}^{z_0}dz\,\frac{h}{z^2}\sqrt{1-\frac{h_0^2}{h^2}\frac{z^4}{z_0^4}}\right]=\\
  \frac{1}{2\pi\alpha'}\left[\sqrt{\text{G}_{00}(z_0)\text{G}_{yy}(z_0)}\,\,r+
  2\int\displaylimits_{0}^{z_0}dz\,\sqrt{\frac{\text{G}_{zz}}{\text{G}_{yy}}}\,\sqrt{\text{G}_{00}\text{G}_{yy}-\text{G}_{00}(z_0)\text{G}_{yy}(z_0)}\right],
\end{multline}
where $\text{G}_{aa}(z_0)=\left.\text{G}_{aa}\right|_{z=z_0}$ for the (Euclidean) time-like, $a=0$,
and space-like, $a=1,2,3$, components, with the latter having non-zero component only along the direction $y$
chosen in Fig.~\ref{wilson_loop}, we denoted this component as $\text{G}_{yy}$.
The last line in~\eqref{E2} follows from a general analysis performed in~\cite{Kinar:1998vq}.
The expression~\eqref{E2} directly shows that the energy of the configuration is the length of the string
(up to a multiplier) according to the metric~\eqref{metric} plus a correction.

Given a specific ansatz for $h(z)$, one can find numerically the function $E(r)$ from the
relations~\eqref{x_z_int} and~\eqref{E1} if these relations are well defined. If $h(z)$ corresponds to a confining
geometry (see below) then asymptotically, at small and large distances $r$, the static energy $E(r)$ will have
the form of the Cornell potential~\eqref{cornell}. The corresponding calculations within several specific SW models
can be found in~\cite{Andreev:2006ct,Afonin:2021zdu,Afonin:2022aqt}. If the geometry of holographic model is non-confining
then the potential will have a Coulomb-like form.

Some comments are in order. The integral in~\eqref{x_z_int} and~\eqref{E1} must be real to
have a physical meaning, i.e., the expression under the square root must be positive.
This condition leads to the emergence of an upper bound on the maximum value of $z$,
\begin{equation}
\label{upper}
  z_0<\frac{2h_0}{h_0'}.
\end{equation}
Here $h_0$ is defined by~\eqref{not0} and $h_0'=\left.\pz h\right|_{z=z_0}$.
The integral develops a logarithmic singularity at $z_0=2h_0/h_0'$ and becomes complex for larger $z_0$.
The appearance of upper bound for a coordinate means the existence of a horizon
which is a generic feature of confining theories.

A simple check shows that the upper bound~\eqref{upper} delivers minimum to the time-time component of
metric~\eqref{metric} given by the condition
\begin{equation}
\label{sonn_cond}
  \left.\pz \text{G}_{00}\right|_{z=z_0}=0,\qquad
  \text{G}_{00}(z_0)\ne0.
\end{equation}
This condition was first derived in~\cite{Kinar:1998vq} as a sufficient condition for confining behavior in the
string theory of the Nambu-Goto type. The last line in~\eqref{E2} also shows that the effective string tension
reaches its minimum at $z = z_0$ since $\text{G}_{yy}=\text{G}_{00}$.

It is clear that the minimum value $z_0$ can emerge only if $h(z)$ in the metric~\eqref{metric} represents
an increasing function of $z$ at large enough $z$, at least in the Euclidean space. For instance, in the SW ansatz
$h(z)= e^{cz^2}$ of paper~\cite{Andreev:2006ct}, this fixes the sign of mass parameter: $c>0$.

Following the arguments in~\cite{br3}, one can provide a nice physical interpretation for the
condition~\eqref{sonn_cond}. In a curved space with confining geometry, a particle falls into
the absolute minimum of the corresponding gravitational potential energy $U$. For a body of mass $m$,
this energy is given by the known relation in general relativity, $U=mc^2\sqrt{G_{00}}$.
The condition~\eqref{sonn_cond} then becomes simply the condition on the minimum of $U$.
Thus, if the potential $U(z)$ has an absolute minimum at some $z_0$ then a particle is confined within distances $z\sim z_0$
and one can imagine a particle confined effectively in a hadron of size $z_0$. Such a physical picture
emerges in the light-front holographic approach, where the holographic coordinate $z$ is a measure for the
interquark distance in a hadron~\cite{br3}.

\section{Ultraviolet regularization of energy}

The integral for the energy in~\eqref{E1} is divergent at $z = 0$ due to the factor $z^{-2}$ in the metric~\eqref{metric}.
This divergence arises universally for both confining and non-confining geometries because their structure at $z\rightarrow0$
should be identical in the holographic models, this structure is completely governed by the AdS$_5$ metric at small $z$
in~\eqref{az_metric}. It cannot be violated by the correction $h(z)$ in sensible holographic models because otherwise
one cannot use the standard holographic dictionary from the AdS/CFT correspondence. The normalization $h(0)=1$
(in general, it can be any constant) was imposed in~\eqref{h} precisely for this reason. The universal structure
of the metric of AdS$_5$ space at small $z$ in~\eqref{metric} leads to a universal Coulomb behavior of static energy
at short distances in the ordinary space with the correct negative sign (which appears in the Hamiltonian~\eqref{ham}).
This Coulomb behavior is a direct consequence of conformal invariance at very short distances, in this way the given
behavior of potential emerges in any conformal field theory, for instance, it was derived for the $\mathcal{N}=4$
SYM theory already in the pioneering paper~\cite{Maldacena:1998im}.
The corresponding calculations within the SW holographic models with a confining geometry were carried out, e.g.,
in~\cite{Andreev:2006ct,Afonin:2022aqt}, they reproduced a general structure of the Cornell potential~\eqref{cornell}.

Usually, in the procedure of eliminating divergence at $z = 0$ in~\eqref{E1}, one introduces the notion of regularized energy.
But for a general discussion of regularization, it will be easier to use the notion of subtracted energy which is finite
and can be directly identified with physical static energy.

The singularities at $z = 0$ in~\eqref{E1} stem from the factor $h(z)/z^2$. In the case of our normalization $h(0)=1$,
the series expansion of $h(z)$ at small $z$ is
\begin{equation}
\label{ser}
\left. h(z)\right|_{z\rightarrow0}=1+\sum_{k=1}^{\infty}c_kz^{\alpha_k},\qquad \alpha_k>0.
\end{equation}
In general situation, this expansion can contain fractional powers of $z$, i.e., $\alpha_k$ can be non-integer.
Assume that there exist $n$ contributions with the values of $\alpha_k$ in the interval
\begin{equation}
\label{alpha}
0<\alpha_k\leq1,\qquad k=1,2,\dots,n.
\end{equation}
Then the integral in~\eqref{E1} will have $n+1$ different divergences, thus we will need $n+1$ subtractions
to regulate the arising ultraviolet singularities. Let us impose the ultraviolet cutoff $\varepsilon$ and
introduce the following regularizing function,
\begin{equation}
\label{regf}
F(\varepsilon,z_0) = \int\displaylimits_{\varepsilon}^{z_0}\frac{dz}{z^2} \left(1+\sum_{k=1}^{n}c_kz^{\alpha_k}\right).
\end{equation}
If $\alpha_n$ is the largest constant in~\eqref{alpha}, the $\varepsilon$-expansion of $F(\varepsilon,z_0)$ has two possible forms
\begin{align}
\alpha_n<1:&\qquad F(\varepsilon,z_0) = \frac{1}{\varepsilon}+\sum_{k=1}^{n}\frac{c_k}{1-\alpha_k}\frac{1}{\varepsilon^{1-\alpha_k}}+\text{const},\\
\alpha_n=1:&\qquad F(\varepsilon,z_0) = \frac{1}{\varepsilon}+\sum_{k=1}^{n-1}\frac{c_k}{1-\alpha_k}\frac{1}{\varepsilon^{1-\alpha_k}}+c_n\ln\frac{z_0}{\varepsilon}+\text{const}.
\label{sec}
\end{align}
Now one can define the subtracted energy as follows,
\begin{equation}
\label{sub}
  E_\text{sub}=\lim_{\varepsilon\rightarrow0}\frac{R^2}{\pi\alpha'}\left(\int\displaylimits_{\varepsilon}^{z_0}dz\,\frac{h}{z^2}
  \frac{1}{\sqrt{1-\frac{h_0^2}{h^2}\frac{z^4}{z_0^4}}}-F(\varepsilon,z_0)\right).
\end{equation}
The energy $E_\text{sub}$ is finite and can be thus identified with the physical energy
between two static sources which is typically measured in the lattice simulations at large enough
separations of sources.

Many calculations of static energy between heavy sources performed within the SW holographic approach exploited regularizations
which are special cases of our general renormalization prescription~\eqref{sub}. For example, in the standard SW model of Ref.~\cite{Andreev:2006ct}
and in its generalizations considered in~\cite{Afonin:2022aqt}, only one subtraction was needed because
in both cases one had $\alpha_1=2$ in the series expansion~\eqref{ser}. In the SW model with linear exponential correction,
$h(z)= e^{\tilde{c}z}$, in the framework of which the heavy-quark potential has been recently analyzed in~\cite{lindil}
(a brief summary of the results is presented in~\cite{qfthepc}), one has $\alpha_1=1$ in the series expansion~\eqref{ser}, thus two
subtractions were required, with the second one being a logarithmic divergence displayed in~\eqref{sec}. In principle, SW holographic
models with arbitrary number of subtractions in static energy can be constructed.

It is clear that other subtraction schemes are possible, they lead to a constant shift
of the finite static energy. If the potential is measured at large separations, the given shift has no direct physical meaning.
However, at short distances, a constant contribution to the potential can carry important physical information.
For instance, in the phenomenological potential models for heavy quarkonia, the constant $C$ in~\eqref{cornell} is
roughly $C\approx-0.3$~GeV~\cite{bali}. The problem of choosing a physical subtraction scheme becomes thus relevant.
This problem is actually open and we will briefly discuss some alternative subtraction schemes.

The first obvious possibility is to change the normalization point in the regularizing function~\eqref{regf},
namely to consider $F(\varepsilon,z_1)$, where $z_1<z_0$. However, as was advocated in~\cite{Brandhuber:1998bs},
various arguments in string theory indicate unambiguously that the subtraction must be at the horizon,
whose role in our case is played by $z_0$.

A more interesting subtraction scheme arises if we redefine the regularizing function~\eqref{regf} in the following way,
\begin{equation}
\label{regf2}
\tilde{F}(\varepsilon,z_0) = \int\displaylimits_{\varepsilon}^{z_0}\frac{h}{z^2}dz.
\end{equation}
The numerator of integrand in~\eqref{regf} contains only the part of $h(z)$ that leads to singularities in energy,
this part is replaced in~\eqref{regf2} by the function $h(z)$ itself. Such a subtraction scheme is completely
equivalent to the scheme proposed by Maldacena in his original paper on the holographic Wilson loop~\cite{Maldacena:1998im},
where it was noticed that the calculated energy includes the masses of infinitely heavy quark\footnote{Actually
the term ``quark'' meant in~\cite{Maldacena:1998im} an infinitely massive vector boson connecting the $N$ branes with one brane
which was far away in the direction $y$. Only in this case the subtraction of mass is well motivated because it appears in
the Wilson loop. We prefer to use the term ``heavy source'' implicitly meaning ``heavy quark''.} and antiquark, therefore the infinity in the
energy is simply due to the fact that the quark masses were included. In the setting under consideration, the ``bare''
quark (i.e., the free quark when the coupling to gauge theory is switched off) should represent a straight string with
a constant value of $y$, stretching from \(z(\pm r/2)=0\) to $z=z_0$~\cite{Maldacena:1998im,Kinar:1998vq}.
Its mass, then, follows from the Nambu-Goto action~\eqref{ng_action},
\begin{equation}
\label{quark}
  m_q=\frac{T}{2\pi\alpha'}\int\displaylimits_{-r/2}^{r/2}dy\,\sqrt{\text{G}_{00}\text{G}_{zz}\,z'^2}
  =\frac{TR^2}{2\pi\alpha'}\int\displaylimits_{0}^{z_0}\frac{h}{z^2}dz,
\end{equation}
where the first term under the square root of the action~\eqref{ng_action} disappears since $y$ is a constant.
Comparing~\eqref{regf2} and~\eqref{quark} one can see immediately that the subtraction of masses of two quarks from the
static energy, $E_\text{sub}=E-2m_q$, is identical to the subtraction of the regularizing function~\eqref{regf2} in~\eqref{sub}.
The given subtraction scheme was widely used in the relevant literature (see, e.g.,~\cite{heavy,Kinar:1998vq}).

\section{Summary}

The Wilson loop confinement criterion provides a nice theoretical way for analytical derivation of the potential between two static sources
which can be measured in the lattice simulations. This gives an interesting opportunity to test various phenomenological holographic
models for strong interactions. In the process of holographic derivation, one inevitably encounters infinity stemming from the
region of vanishing values of holographic coordinate. In the case of Soft-Wall holographic models, many types of such infinities become possible,
although in practice the appearance of more than two types looks exotic. The model with linear exponential background in the function $h(z)$
of modified AdS metric, see~\eqref{az_metric}, gives an example when two different divergences in the potential energy arise. If the series
expansion of $h(z)$ at small $z$  contains additional fractional powers $z^\alpha$ in the interval $0<\alpha<1$, then those additional exotic
divergences arise. Such a situation is not excluded both in the phenomenological static holographic models and in dynamical dilaton-gravity
holographic systems. Various subtraction schemes can be applied to regulate the emerging divergences, they result in a different constant term
in the potential energy and this contribution becomes thus scheme-dependent. We hope that the clarification of the raised issue will be helpful
for further applications of bottom-up holographic models to the physics of confinement and related phenomenology.

\section*{Acknowledgements}

This research was funded by the Russian Science Foundation grant number 21-12-00020.



\end{document}